# Solid-Immersion Metalenses for Infrared Focal Plane Arrays

Shuyan Zhang[1, #], Alexander Soibel[2,*], Sam A. Keo[2], Daniel Wilson[2], Sir. B. Rafol[2], David Z. Ting[2], Alan She[1], Sarath D. Gunapala[2], and Federico Capasso[1,*]

[1]John A. Paulson School of Engineering and Applied Sciences, Harvard University, 9 Oxford Street, Cambridge, MA 02138
[2]Jet Propulsion Laboratory, California Institute of Technology, 4800 Oak Grove Dr., Pasadena, CA 91011
[#]Present address: Singapore Bioimaging Consortium, A*STAR, 11 Biopolis Way, Singapore 138667
[*]Email: asoibel@jpl.nasa.gov, capasso@seas.harvard.edu

## Abstract

Novel optical components based on metasurfaces (metalenses) offer a new methodology for microlens arrays. In particular, metalens arrays have the potential of being monolithically integrated with infrared focal plane arrays (IR FPAs) to increase the operating temperature and sensitivity of the latter. In this work, we demonstrate a new type of transmissive metalens that focuses the incident light ($\lambda = 3 - 5$ μm) on the detector plane after propagating through the substrate, i.e. solid-immersion type of focusing. The metalens is fabricated by etching the backside of the detector substrate material (GaSb here) making this approach compatible with the architecture of back-illuminated FPAs. In addition, our designs work for all incident polarizations. We fabricate a 10x10 metalens array that proves the scalability of this approach for FPAs. In the future, these solid-immersion metalenses arrays will be monolithically integrated with IR FPAs.

Infrared focal plane arrays (IR FPAs) are commonly used in thermal cameras, medical imaging devices and for sensing applications such as wavefront sensing[1]. Microlenses and microspheres have been previously used as optical concentrators[2-9] to increase the operating temperature of IR FPAs, but they are typically made of materials different from the detector materials and therefore these approaches require additional deposition and alignments steps. Recently, mid-wavelength IR (MWIR) nBn detectors monolithically integrated with spherical concentrators fabricated on the detector backside were demonstrated[10]. This provides an alternative approach for realization of microlens-integrated detectors in which microlenses are made from the detector substrate material. The newly developed metasurfaces are a promising candidate for the next generation optical concentrators that can also be monolithically integrated with IR FPAs with small pixels. They can be fabricated from the same material as the substrate and are flat, ultrathin, and lightweight. Metasurfaces consist of optical components based on arrays of optical resonators with subwavelength separation. By accurately designing the optical properties of each element of the array, the wavefront of the incident light can be reshaped and redirected at will[11]. Numerous devices based on metasurfaces have been developed, including metasurface lenses (metalenses), waveplates, polarimeters, and holograms[12-19].

To be compatible with current IR FPAs, these metalenses have to feature several unique characteristics which differentiate them from the metalenses demonstrated so far. The majority of IR FPAs are back-illuminated (through the substrate), so the lens needs to be transmissive and of the immersion type to focus light in the detector materials. The lens also needs to be fabricated on the backside of the detector so it is made of the same material as the detector wafer substrate (GaSb, InSb or CdTe). Hence, the fabrication process needs to be compatible with the detector manufacturing. Furthermore, it has to be scalable to the size of microlens arrays and robust to avoid issues with FPA uniformity and operability. Finally, similar to many other applications, the lens should have a high focusing efficiency, work for all incident polarizations and be broadband (e. g. 3 – 5 µm or 8 – 12 µm).

In this work, we demonstrate a new type of transmissive metalens that addresses many of these challenges. The schematic is shown in Figure 1a. It focuses light from air into the substrate material, i.e. solid-immersion as compared to in air in normal cases, with a focusing efficiency up to 52% in experiment and 80% in simulation (Figure 1b). The metalens consists of circularly shaped posts etched directly into a GaSb substrate. The circular shape of the posts ensures that the designs work for all incident light polarizations. By varying the diameter of the posts, a phase coverage of 2π and a relatively high and uniform transmission amplitude response can be achieved (see details in Supplementary Information Section 1). The phase profiles of the metalenses were realized with these subwavelength posts with fixed edge-to-edge separation, by which the placement of posts was made denser than that of metalenses with fixed center-to-center separation[8]. We fabricated a 10x10 metalens array with each metalens having a 30 µm in diameter and a focal spot size of 10 µm (typical FPA pixel pitch size and pixel size), which to our knowledge is the smallest size optical concentrator for FPAs demonstrated so far. In particular, we demonstrated a "chromatic metalens", i.e. metalens not corrected for chromatic aberration, operating at $\lambda$ = 4 µm and a "broadband metalens" with achromatic behavior at 3 – 4 µm. We have also studied their optical performance in the entire MWIR wavelength range, 3 – 5 µm. In the future, the solid-immersion metalens will be monolithically integrated with the IR FPAs as shown in Figure 1c. Each pixel in the FPA will have its own metalens that acts as an optical concentrator to enhance the light collection. We will use the semiconductor wafer where the active detector material is grown on the top GaSb substrate. We will fabricate the metalens arrays on the back side and the detector array on the front side using existing processes reported in[20]. Next, the integrated metalens-detector array will be hybridized with Si read-out circuits into FPAs. Our results open a path to the realization of IR detectors and FPAs integrated with a metasurface-based focusing lens with improved signal-to-noise ratio due to a reduced detector volume for various sensing and imaging applications.

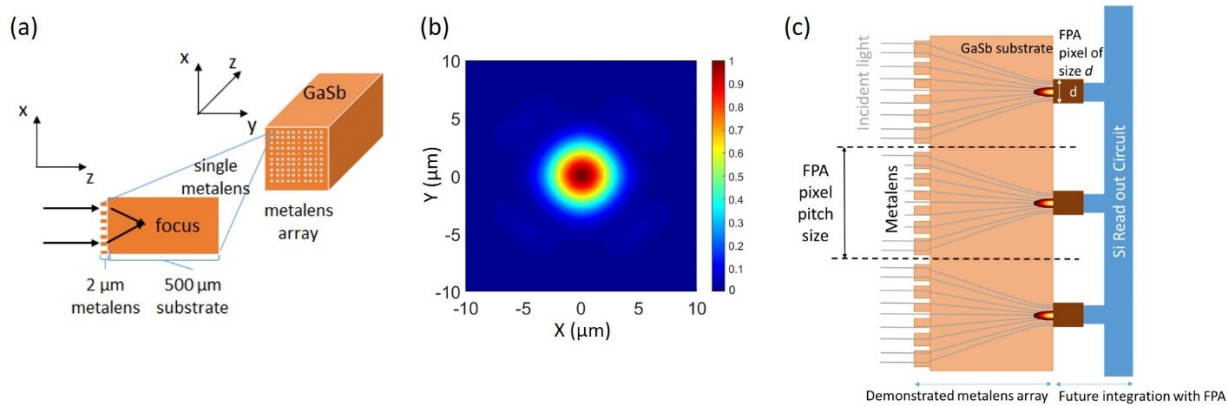

**Figure 1:** (a) Solid-immersion metalens schematic: (Right) Metalens array is on the backside of the substrate. (Left) Zoom-in cross-sectional view of a single metalens (2-µm-thick) showing that the incident

light is focused into the substrate (500-µm-thick). (b) Simulated electric field intensity (normalized) in the *x-y* plane at the focus. The simulations were performed using the FDTD software package from Lumerical, Inc. (c) Proposed future work of integration of a solid-immersion metalens with an FPA pixel unit cell.

For fabrication of the metalenses, the substrates were first prepared by solvent clean (IPA and acetone), oxygen plasma ash, and buffered oxide etch. Then a hard mask made of $SiN_x$ was deposited on the wafer. The metalenses were patterned by electron-beam lithography using a ZEP520A resist with a film thickness of around 500 nm. After the resist development, the hard mask was etched with $CF_4$ and $O_2$ using an inductively-coupled plasma (ICP) system (Plasma-Therm ICP etching system). Prior to the metalens etching, samples were mounted on substrate carriers with a thermal conductive-cooled adhesive medium. The GaSb was etched with a mixture of $Cl_2$, $BCl_3$, and Ar in an ICP etching system. The hard mask was then removed. Figure 2a shows the scanning electron microscopy (SEM) images of the fabricated metalens array. Figure 2b is a zoom-in view of a single metalens. Figure 2c shows individual posts with smooth sidewalls after etching. Because of the high aspect ratio of the posts and the fact that GaSb is difficult to etch, we obtained a sidewall angle of approximately 83°. The corresponding phase and transmission amplitude response are discussed in the Supplementary Information Section 1.

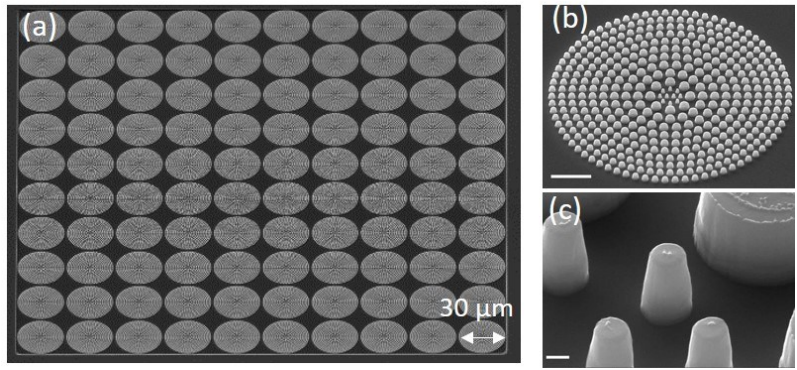

**Figure 2:** (a) SEM image of 10x10 metalens array. Each metalens is 30 µm in diameter. (b) SEM image of a single metalens. The individual posts comprising the lens are clearly visible. Scale bar: 5 µm. (c) SEM image of individual posts (SEM image tilting angle is 50°). Scale bar: 200 nm.

The experimental setup is shown in Figure 3a. The blackbody at 1273 K was used as the infrared source which has a broad emission spectrum. The exit aperture diameter of the blackbody is 1.5 cm. Initially, we attempted to investigate spectral response of the metalenses using grating spectrometer but the signals were too low. Therefore, we performed measurements with five bandpass filters: BP-3050-200 nm (2950 - 3150 nm), BP-3390-345 nm (3220 - 3565 nm), BP-3900-200 nm (3800 - 4000 nm), BP-4220-200 nm (4120 - 4320 nm), and BP-4665-240 nm (4545 - 4785 nm). The metalenses were placed after the bandpass filter with the posts facing the bandpass filter. Although the output beam of the blackbody source is not collimated, the distance from the blackbody source aperture to the metalens array is about 20 cm, we could approximate that the incident light hitting a single metalens is collimated and at normal incidence. Imaging was performed with a 36X reflective microscope objective (Newport, 50102-01) and an MWIR infrared camera (1024x1024 pixels, JPL developed). The infrared camera with attached microscope objective was mounted on a translational stage. During the measurement, the camera assembly was moved along the optical axis (dashed line), i.e. *z*-direction to image inside the substrate. Figure 3b and 3c show the acquired IR images of the metalens array taken at different $z$ values after subtracting the background intensity taken when the objective was blocked. $Z = 0$ µm was set when the camera was focused on the metalens (Figure 3b). The metalens shapes are clearly visible. The camera assembly was then moved away from the metalens in 1 µm steps. There is a subtle point in this measurement that is important to clarify. The measured $z$ values of the objective movement in air need to be multiplied by the refractive index of GaSb. This is done to find the position of focus within GaSb (see Supplementary Information Section 2). Figure 3c is the camera image taken at the focal plane of the metalens array.

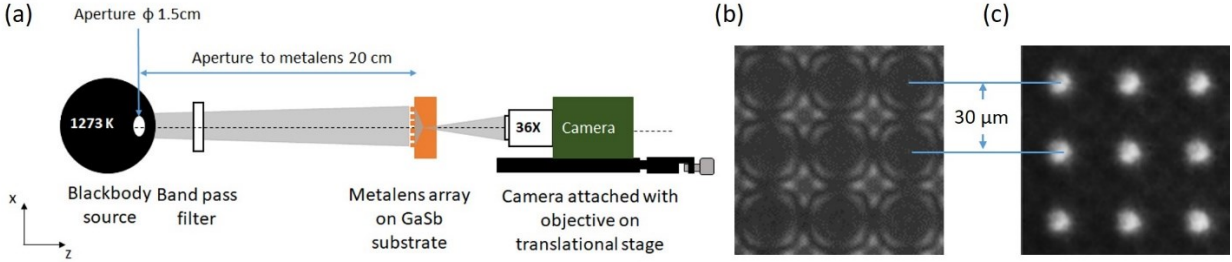

**Figure 3**: (a) Experimental setup. The metalens array was facing towards the bandpass filter. (b) Camera image of a 3 x 3 subsection of metalens array when the camera was focused on the metalenses. The center-to-center distance of the neighboring metalenses is 30 µm. (c) Camera image at the focal plane showing the intensity enhancement. (b) and (c) were obtained using the broadband metalens array with the bandpass filter BP-3390-345 nm.

The measurement results of the two metalens designs are shown in Figure 4. The measured normalized intensity in the *x-z* plane is shown in Figure 4a and 4c for the chromatic metalens and broadband metalens, respectively. The beam profiles at the focus are shown in Figure 4b and 4d, which are the line cuts at the maximum intensity values of Figure 4a and 4c. The dots are measured data and the lines are Gaussian fits with an $R^2 > 0.95$. The $x$ values are relative, so they are shifted so that the peaks align at $x = 0$. For the chromatic metalens, the measured full beam waist (FBW) at $1/e^2$ of the peak intensity is 13.9 µm at $\lambda = 3.9$ µm. For the broadband metalens, the average FBW is 16.4 µm for $\lambda = 3 - 5$ µm. The corresponding Strehl ratios are 0.52 and 0.38 calculated according to the definition given in[21]. The measured FBW values are greater than the diffraction-limited spot size is possibly caused by the incident light not being single wavelength and normal incidence and also fabrication imperfections, e.g. the slope of the fabricated posts is not 90° as designed. Note that it is not necessary to achieve diffraction-limited focusing as long as the focused spot size is smaller than the FPA pixel size because the focused beam profile is not important. What is important is the intensity at the focus. In addition, the focused spot size of the metalens can be adjusted by changing the NA of the design to suit different FPA pixel sizes. For example, further reduction of the focused beam size to below 10 µm is achievable by increasing the numerical aperture (NA). The NAs of the current chromatic and broadband metalenses are 0.36 and 0.35, respectively. Figure 4e shows the measured focal length as a function of the incident wavelength for the two designs. For the chromatic metalens, the focal length at $\lambda = 3.9$ µm is 149.2 µm which is close to the designed value ($f = 155$ µm). The focal length decreases as the incident wavelength increases. The focal length of the broadband metalens is constant for the first three bandpass filters, i.e. 152.9 µm for $\lambda = 2.95 - 4$ µm. The designed value is $f = 158$ µm. As the wavelength increases further, the focal length decreases, similar to the chromatic metalens. The simulation results of the broadband metalens are provided in the Supplementary Information

Section 3. Both our simulation and measurement results show that the current broadband metalens design is only broadband from 3 – 4 μm, however, the percentage change is relatively small $\Delta f/f$ = 14.6% for λ > 4 μm. There have been recent demonstrations showing ways to simultaneously control the phase, group delay and group delay dispersion of light to achieve a transmissive achromatic metalens with larger bandwidth[22, 23].

Responsivity is an important parameter of FPAs. Metalenses act as optical concentrators to collect incident light and focus it onto the active area (pixel size) of the FPA pixel unit cell (pixel pitch size) after propagating through the substrate material, hence the device responsivity is increased. To quantify the focusing power of the metalenses, we measured the focusing efficiency and the intensity enhancement of the metalens as an optical concentrator. The focusing efficiency was defined as the optical power over the pixel size of 10 μm at the focus divided by the incident power over the pixel pitch size of 30 μm. The measured values are plotted in Figure 4f for different bandpass filters for the broadband metalens. The maximum focusing efficiency measured is 52% with the bandpass filter BP-3390-345 nm. In the simulation, the average focusing efficiency is 70% over a wavelength range of 3 to 5 μm and the maximum focusing efficiency is 80%. The loss mainly comes from the reflection at the GaSb/air interface. The focusing efficiency can be improved by matching the optical impedance of the planar device with that of free space[24] and/or depositing a layer of anti-reflection coating. The intensity enhancement is defined as the ratio between the peak intensity at the focus and the incident intensity. We observed an intensity enhancement of approximately 3 times, i.e. 300%, for both the chromatic metalens and broadband metalens.

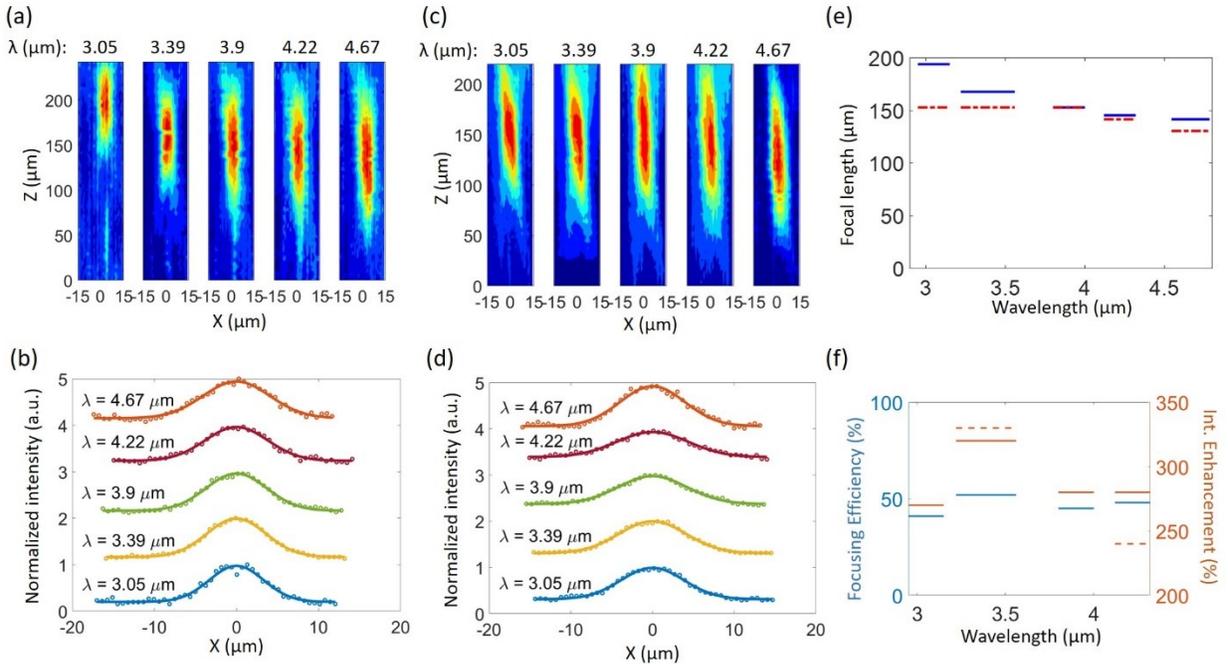

**Figure 4**: Measurement results for the chromatic metalens and broadband metalens: (a) and (c) Normalized intensity in the *x-z* plane for different incident wavelengths. (b) and (d) Beam profiles at the focus for (a) and (c), respectively. The wavelength labels in (a), (b), (c), and (d) are taken as the center wavelengths of the bandpass filters. (e) Focal length of the chromatic metalens (blue) and broadband metalens (red) as a function of the incident wavelength. (f) Measured focusing efficiency (blue solid line) and intensity

enhancement (orange solid line) of the broadband metalens and intensity enhancement (orange dashed line) of the chromatic metalens for different incident wavelengths with different bandpass filters.

In summary, we successfully demonstrated a new type of metalens that operates as a solid-immersion lens, i.e. incident light is focused into the substrate material. The metalenses comprising of GaSb posts were fabricated directly on the backside of GaSb substrates making it compatible with backside illuminated IR FPAs. To keep the detector volume small, our metalenses are 30 μm in diameter corresponding to the FPA pixel pitch size and only 2 μm thick which are much smaller than the typical size of microlenses and microspheres for FPAs. Two metalens designs were studied. The chromatic metalens operating at $\lambda = 4$ μm has a smaller focal spot size as compared to the broadband metalens. But the broadband metalens has a constant focal length at $\lambda = 3 – 4$ μm and has a smaller focal length variation over the incident wavelength range (3 – 5 μm). Both designs show intensity enhancement of around three times at the focus indicating a potential to improve responsivity of the FPAs. In particular, the demonstrated 10x10 metalens array proves the scalability of this approach for FPAs. Our demonstration opens a path to the realization of high operating temperature IR detectors and FPAs monolithically integrated with flat and lightweight metalenses.

The authors thank Xinghui Yin for helpful discussions and A. Shapiro-Scharlotta, F. Y. Hadaegh and R. Howard for encouragement and support. This work was supported by the Center Innovation Fund, Space Technology Mission Directorate, National Aeronautics and Space Administration. Shuyan Zhang thanks A*STAR Singapore for support through the National Science Scholarship. Part of this research was carried out at the Jet Propulsion Laboratory, California Institute of Technology, under a contract with the National Aeronautics and Space Administration.

# Supplementary Information

1. Phase and transmission amplitude of GaSb posts

Figure S1 shows that by varying the diameter of the posts, a phase coverage of $2\pi$ and a relatively high and uniform transmission amplitude response can be achieved. The data points for the GaSb posts with 90º sidewall angle is shown as circle markers. The dip of the transmission amplitude is due to the electric dipole resonance. The phase and amplitude data were used as a lookup table for digitizing the phase profile of the metalenses. The phase profiles of the metalenses were realized with these subwavelength posts with fixed edge-to-edge separation, by which the placement of posts was made denser than that of metalenses with fixed center-to-center separation. In our design, the edge-to-edge spacing was 400 nm, which was chosen to minimize optical coupling between neighboring posts. Note that only posts with a diameter range from 440 to 1160 nm (for the chromatic metalens) and from 740 to 980 nm (for the broadband metalens) were necessary to achieve the phase coverage required.

The fabricated posts have a sidewall angle of 83º instead of the designed 90º. We have simulated the effect on the phase and transmission amplitude (star markers). The curves are shifted towards the direction of bigger post diameters. For the transmission amplitude, there is no electric dipole resonance dip. For the phase response, the biggest deviation between the 90º and 83º is $\Delta\varphi = 0.39\pi$ for the chromatic metalens and $\Delta\varphi = 0.26\pi$ for the broadband metalens. This phase error introduced by the fabrication contributes to a degree of blurring, resulting in a measured focused spot size larger than the diffraction-limited spot size.

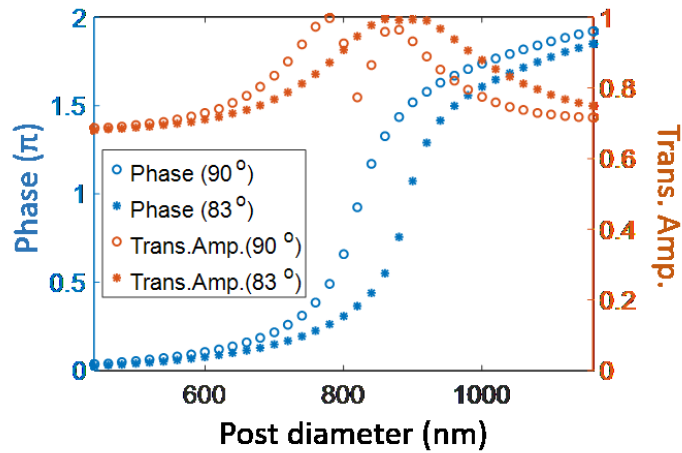

**Figure S1:** Simulated phase and transmission amplitude response of posts with 90º sidewall angle (circle markers) and 83º sidewall angle (star markers).

## 2. Calculations of the measured z values

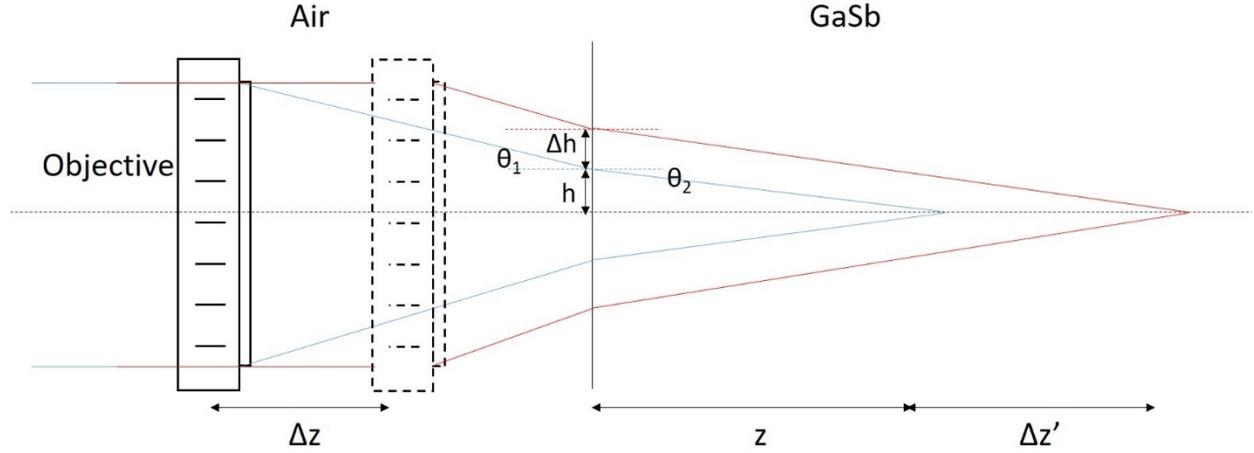

**Figure S2**: The objective was moved along the optical axis by a distance $\Delta z$ in the air medium. We are interested in what the actual movement is in the GaSb medium, i.e. $\Delta z'$.

Based on the Snell's Law, we have:

$$n_{Air} \times sin(\theta_1) = n_{GaSb} \times sin(\theta_2)$$

For small angles ($\theta_1 = 5.5°$ in this case):

$$\theta_1 = n_{GaSb} \times \theta_2$$

Apply small angle approximation:

$$\Delta h = \Delta z \times tan(\theta_1) = \Delta z \times \theta_1$$

In GaSb:

$$h = z \times tan(\theta_2)$$

$$\Delta h = \Delta z' \times tan(\theta_2) = \Delta z' \times \theta_2$$

Hence, we have:

$$\Delta z \times \theta_1 = \Delta z' \times \theta_2$$

$$\Delta z' = \Delta z \times \frac{\theta_1}{\theta_2} = \Delta z \times n_{GaSb}$$

## 3. Simulation results of the broadband metalens

For the broadband metalens, the incident wavelength range is from 3 to 5 µm. The post height is 2 µm throughout and the post diameter range is 740 – 980 nm which is photolithography compatible. The post height-to-diameter ratio is smaller than 2.7 to make the etch process within the process bounds. Figure S3a is the beam size as a function of the incident wavelength and the $x$ position at the focus $z = f = 158$ µm. The two white dashed lines indicate the pixel size of the FPA detector, i.e. 10 µm. The beam size is confined within the 10 µm window for the entire wavelength range. Figure S3b plots the focused beam profile of selected wavelengths. We can see that the beam size stays relatively constant. Figure S3c shows the focal length as a function of wavelength. For achromatic focusing, a flat curve is desired. The black dashed line indicates the designed focal length. The lens is therefore achromatic for $\lambda = 3.2 – 4.1$ µm, i.e. the achromatic bandwidth is $\Delta\lambda = 0.9$ µm, consistent with the measured results. The uncertainty of the focal length is 2 µm, which is the resolution in the $z$-axis in the simulation.

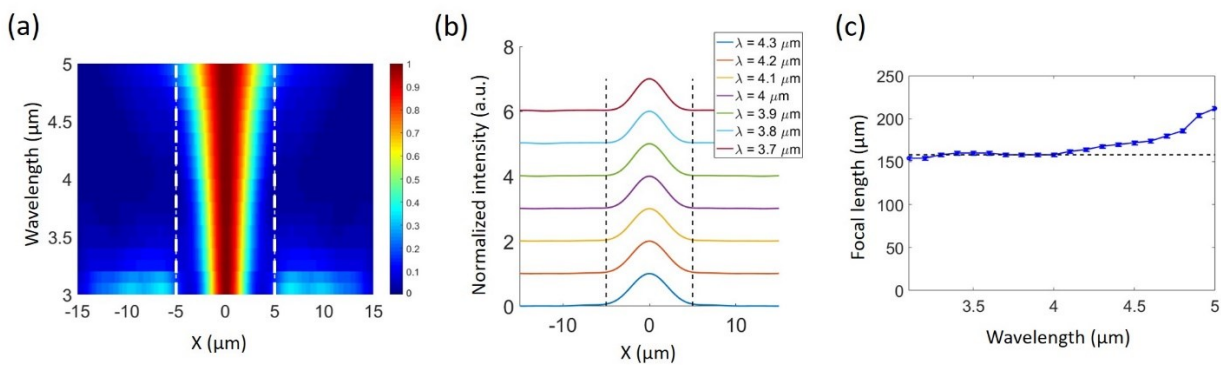

**Figure S3:** FDTD simulations of the broadband metalens. (a) Beam profile as a function of incident wavelength and $x$ position at the focus. (b) Horizontal line cuts of (a) at selected wavelengths. Curves are shifted in the $y$-axis for visual clarity. (c) Focal length as a function of wavelength. The black dashed line indicates the designed focal length, 158 µm.